# Title: Direct observation of topological magnon edge states

Jihai Zhang[1,2]†, Meng-Han Zhang[1,2]†, Peigen Li[1,2]†, Zizhao Liu[1,2], Ye Tao[1,2], Hongkun Wang[1,2], Dao-Xin Yao[1,2]*, Donghui Guo[1], and Dingyong Zhong[1,2]*

[1]School of Physics & Guangdong Provincial Key Laboratory of Magnetoelectric Physics and Devices; Sun Yat-sen University, 510275 Guangzhou, China
[2]State Key Laboratory of Optoelectronic Materials and Technologies, Sun Yat-sen University, 510275 Guangzhou, China
*Corresponding author. Email: dyzhong@mail.sysu.edu.cn (D.Z.); yaodaox@mail.sysu.edu.cn (D.-X.Y.)
†The authors equally contributed to this work.

**Abstract:** Magnon Chern insulators (MCIs) exhibit unique topological magnon band structures featuring chiral edge states. Direct observations of the topologically protected magnon edge states have long been pursued. Here, we report the spatially resolved detection of magnon edge states in a two-dimensional ferromagnet with honeycomb lattice (single-layer chromium triiodide). Using scanning tunneling microscopy, we observed magnon-assisted inelastic tunneling conductance and revealed the gapped magnon spectra with enhanced signals at the van Hove singularities. Extra tunneling conductance contributed from the magnon edge states was detected at three different edge configurations. Our work provided direct evidence proving the existence of MCI states down to the single-layer limit, initiating spatially-resolved explorations on exotic properties arising from topological edge states of MCIs.

**Main Text:**

Magnons are the quanta of spin excitations (spin waves) in magnetic materials. As charge-neutral bosonic quasiparticles, magnons can serve as information carriers with promising applications in ultra-low-power spintronic devices (*1–4*). Magnon Chern insulators (MCIs) feature topological magnon band structures analogous to the quantum Hall states of electronic systems. In MCIs, magnons perform unique transport behaviors undergoing deflections perpendicular to the propagation direction due to the non-zero Berry curvature (*5, 6*), resulting in intriguing phenomena such as magnon Hall effects (*7*). The magnon band structures of MCIs have gapped bulk states with non-zero Chern numbers but topologically protected gapless edge states (*8*), which may be driven by Dzyaloshinskii-Moriya interactions (DMI) breaking the effective time-reversal ($\mathcal{T}$) and inversion ($c_x$) symmetries (*9, 10*). To date, a number of magnetic materials have predicted as MCIs (*6, 11, 12*). Directly observing magnon edge states will not only provide strong evidence supporting the existence of topological magnon states, but also initiate the explorations on exotic properties arising from topological magnon states (*2*). However, due to their localized feature and relatively tiny proportion of density of states (DOS) compared with bulk magnons, it is hard to detect with certain bulk-sensitive techniques such as neutron scattering and Raman spectroscopy. A detection technique, which is sensitive with spin excitations with high spatial and energy resolution, is required.

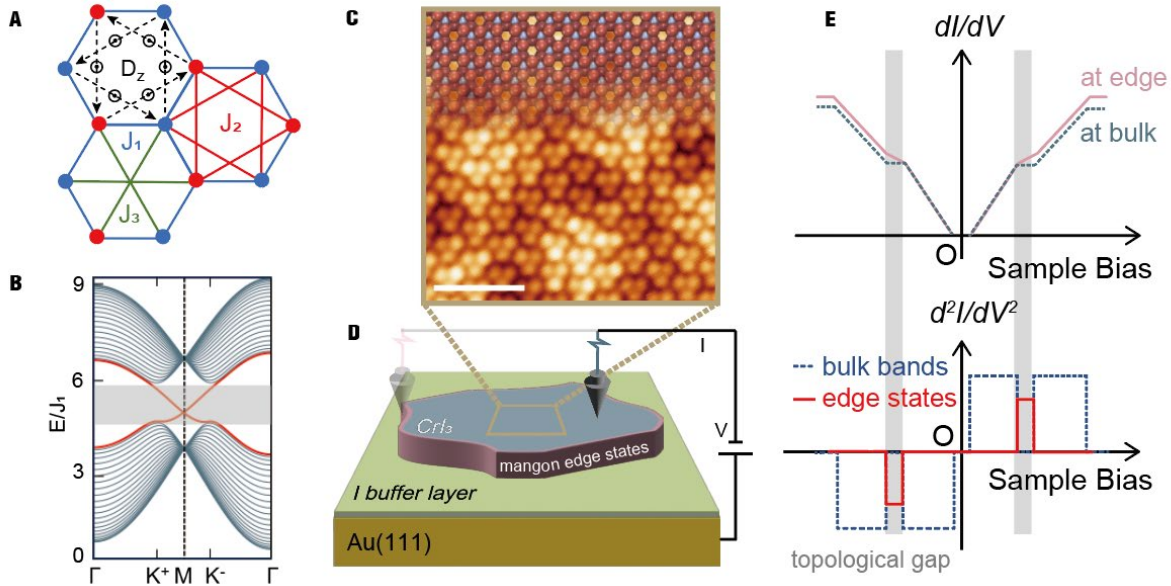

**Fig. 1. Schematics for detecting topological magnon edge states of 2D MCIs using IETS.** (**A**) Ferromagnetic honeycomb lattice with the nearest ($J_1$), next-nearest ($J_2$) and 3rd-nearest neighbor ($J_3$) Heisenberg exchange interactions and Dzyaloshinskii-Moriya interactions with an out-of-plane component ($D_z$). (**B**) Topological magnon bands of an armchair-edged ribbon of the ferromagnetic honeycomb lattice with gapped bulk bands (stone blue) and gapless edges states (red), the topological gap is highlighted by gray shade. (**C**) Atomically resolved STM image of single-layer $CrI_3$ on Au(111) (V = 0.7 V, I = 100 pA). The upper part of the image is superposed with the structural model. Bar, 2.0 nm. (**D**) Experimental scenarios. Single-layer $CrI_3$ is grown on Au(111) substrate with an iodine passivating monolayer. Magon-assisted inelastic tunneling signals are acquired from the interior and edge of the $CrI_3$ island, revealing spatially-resolved magnon spectra. (**E**) Schematics of magnon-assisted inelastic tunneling conductance. In the interior of the islands, the topological magnon gap corresponds to a plateau in the $dI/dV$ channel

and a trough in the $d^2I/dV^2$ channel. At the edge, the edge states are expected to contribute extra signals in $d^2I/dV^2$.

Analogous to the Haldane's quantum anomalous Hall effect model, MCI states may emerge in two-dimensional (2D) ferromagnetic honeycomb lattices (Fig. 1A) (*13*, *14*). The non-trivial topology of the magnon band structure is characterized by the first-class Chern numbers of the optical/acoustical branches of spin waves (*15*),

$$C^{\pm} = \pm \frac{1}{2\pi} \iint_{BZ} dk_x dk_y \, \Omega_k^{\pm},$$

where $\Omega_k^{\pm}$ is the Berry curvature (*16*). Without DMIs, the Dirac magnon band degeneracy located at the $K^{\pm}$ points of the Brillouin zone is protected by the combined $\mathcal{T}c_x$ and six-fold rotation symmetry ($C_{6v}$), with the topological classification belonging to $\mathbb{Z}\oplus\mathbb{Z}$ (*17*). Breaking the $\mathcal{T}c_x$ symmetry by next-nearest neighbor (NNN) DMIs opens a gap at $K^{\pm}$. Due to the topological nature, gapless states spatially localized at the edges of an open-boundary system emerge in the energy gap between the two bulk branches (Fig. 1B). A well-known material system is single-layer chromium triiodide ($CrI_3$) (Fig.1C), which is one of the earliest confirmed 2D ferromagnets with out-of-plane spin polarization in the ground state (*18*, *19*). The honeycomb structure and the strong spin-orbital coupling induced by the iodine atoms make single-layer $CrI_3$ a promising candidate as MCIs (*12*, *20*). By inelastic neutron scattering (INS), a substantial magnon gap of 2.8 meV between the two magnon bands has been observed in multilayer $CrI_3$ crystals (*21*), which is a crucial step suggesting single-layer $CrI_3$ as an ideal platform for further exploring the topological magnon edge states.

Here, using inelastic electron tunneling spectroscopy (IETS) based on scanning tunneling microscopy (STM), we investigated spatially resolved magnon spectra in single-layer $CrI_3$. IETS is a powerful tool to characterize local excitations such as vibrational modes of single molecules (*22*) and spin-flip processes of local spins (*23–26*). In our work, a prominent magnon gap centered at 11.0 meV was identified in the IETS of epitaxial single-layer $CrI_3$. The magnon-related origin of the excitation gap was further confirmed by the Zeeman-like energy shift of the signals induced by an external magnetic field. More importantly, extra inelastic tunneling signals corresponding to the topological magnon edge states were detected within the energy gap at the edges of $CrI_3$. Our observations provide direct evidence for the existence of topological magnon edge states, affirming the bulk-boundary correspondence of MCI states in 2D ferromagnets (*27*).

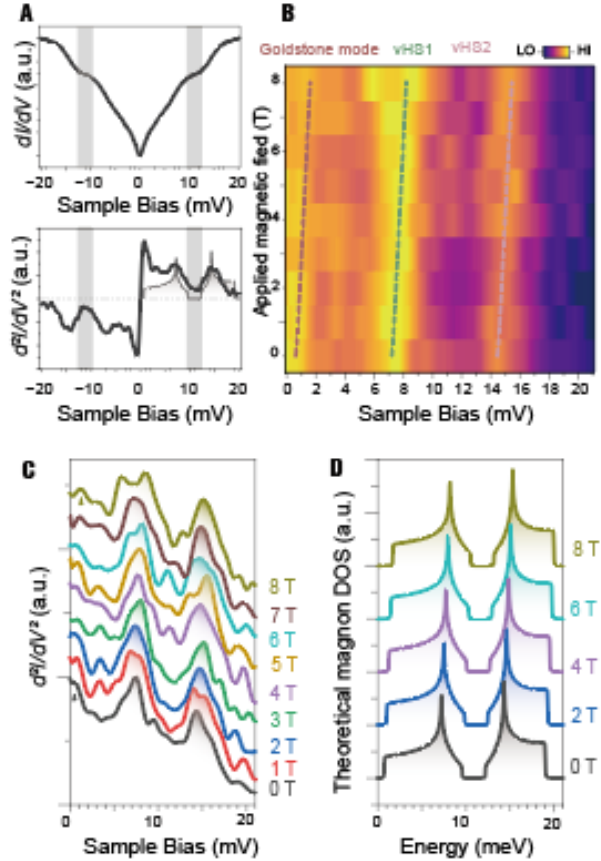

**Fig. 2. Magnetic-field-dependent IETS of single-layer CrI$_3$.** (**A**) Typical IETS measured at the center of CrI$_3$. The upper panel is the symmetrized differential conductance, and the lower panel is the numerical differentiation result based on the smoothed differential conductance. The magnon gap is highlighted by grey shades. (**B, C**) Colored map and curves of IETS under an external magnetic field ($B_\perp$) up to 8 T. The Zeeman-like energy shift verifies the predominate contributions of magnons to the IETS. The dashed lines in B show the simulated energies of the Goldstone gap cliff and vHSs as a function of $B_\perp$. The arrows in C marked the position of the peak related to Goldstone gap. (**D**) Simulated magnon DOS at a finite temperature (0.5 K) based on the field-dependent renormalization.

**Evidence for the magnon gap**

Compared with the IETS induced by discrete excitation modes exhibiting symmetric step-like behaviors in the differential conductance ($dI/dV$) channel and peak-like signals in the second-order differential conductance ($d^2I/dV^2$) channel (*23–25*), the IETS signals for continuous excitations in the $d^2I/dV^2$ channel are approximately proportional to the DOS of the local excitations and the $dI/dV$ channel can be regarded as the integral of DOS (*28, 29*). As illustrated in Fig. 1D, owing to the spatial resolution of STM, it is able to detect the bulk topological magnon gap and obtain extra signals contributed by the edge states within the topological gap by IETS. When the tip is positioned upon the interior region, the $d^2I/dV^2$ channel exhibits two continuous excitations separated by a trough, corresponding to the two bulk magnon branches separated by the topological magnon gap. Consequently, a plateau-like feature appears in $dI/dV$ channel within the gap. At the edges of the system, on the other hand, the existence of the topological magnon edge states is expected to induce additional signals within the topological magnon gap (*30*). As a result, the topological gap will effectively "close" at the edge, and the plateau observed in the *dI/dV* channel will disappear. In other words, the different behaviors at the edges compared with the bulk offer an explicit criterion for the existence of edge states (Fig.1E).

Experiments were conducted on single-layer $CrI_3$ grown on iodine-passivated Au(111) surfaces by molecular beam epitaxy (Fig. 1C and D). In the STM image, there are typical trimerized protrusions assigned to the top layer of iodine superposed by a contrast modulation of moiré pattern owing to the interfacial lattice mismatch (*31*). The insulating nature of both the iodine buffer layer and the $CrI_3$ monolayer (Fig. S1) effectively eliminates the conductance contributed by the elastic resonant tunneling. With a silver-coated tip (Fig. S2), the IETS collected from different positions in the interior of $CrI_3$ islands show similar behaviors and no obvious modulation from the moiré pattern was observed (Fig. S3). Typical IETS results for the interior of $CrI_3$ islands and the $d^2I/dV^2$ curves are shown in Fig. 2A. The *dI/dV* channel approximately has a three-step V-shaped curve with two plateaus, while the $d^2I/dV^2$ channel shows three peaks and two troughs correspondingly. The bias-symmetrical feature of the signal implies that the conductance is contributed by inelastic tunneling. The bias-dependent intensity of $d^2I/dV^2$ signal represents the DOS of magnons integrated from the magnon bands of $CrI_3$ as obtained by INS (*21*).

Similar magnon-assisted IETS was acquired by STM on single-layer $CrBr_3$ grown on graphite (*29*) and by transport measurement of a few-layer $CrI_3$ tunneling device (*32*). One of the significant features of our IETS is the trough (local minimum) of $d^2I/dV^2$ curves with the energy centered at 11.0 meV, at which the intensity is as weak as that with an energy exceeding 19 meV (above the upper limit of the magnon bands). The result confirms the occurrence of magnon gap in $CrI_3$ down to the single-layer limit, analogous to the multilayer crystal (*12*, *21*). Note that the non-zero $d^2I/dV^2$ intensity within the magnon gap may arise from certain excitations such as phonons, which have a finite DOS with the energy below 15 meV and may contribute a featureless background to the IETS (*33*, *34*). Furthermore, due to the broadening effect of IETS, the energy range of the trough in the $d^2I/dV^2$ channel is slightly shrunken compared with the value (2.8 meV) of multilayer crystals. Besides the trough corresponding to the topological gap, another significant feature of our IETS results is the three peaks at 0.7, 7.2, and 14.4 meV, respectively. The last two peaks are attributed to the van Hove singularities (vHSs) in the magnon bands, consistent with the INS results (*21*), while the first one can be assigned to the cliff of the Goldstone gap. According to the Goldstone theorem, the breaking of SU(2) symmetry by anisotropic field in a ferromagnetic spin system generates a zero-energy mode at the $\Gamma$ point, which dominates the high-order corrections in low-energy dynamics and thermodynamics (*35*).

**Evolution of the IETS under vertical magnetic field**

In our magnon-assisted IETS, an energy gap centered at 11.0 meV was clearly observed. To further clarify the physical origin of the IETS signals, an out-of-plane magnetic field ($B_\perp$) up to 8 T was applied on the samples to explore the field-dependent evolution of the magnon spectra. According to the theoretical model, the magnon bands in single-layer $CrI_3$ are expected to blue-shift with increasing $B_\perp$, which strengthens the spin stiffness of the ferromagnetic systems. The experimental results, depicted in the color map of $d^2I/dV^2$ in Fig. 2B, show that all the three peaks shift to higher energies as $B_\perp$ increases, confirming that the inelastic tunneling is dominated by spin excitations. An energy shift about 1.2 meV was obtained with $B_\perp$=8 T, or 0.15 meV/T, which is close to the Zeeman energy (0.12 meV/T) in our system. Furthermore, we found that the energy shift of the magnon spectra exhibits a slightly nonlinear relationship with $B_\perp$, analogous to the result of a few-layer $CrI_3$ tunneling junction device (*32*). In a multilayer sample, interlayer interactions have non-ignorable impact on the magnetic ground states and magnon band structures (*12*, *18*, *21*, *36*). Given the simplicity of the single-layer system without interlayer magnetic coupling, our result reflects the intrinsic properties of a 2D honeycomb ferromagnet. Here, the position of the first peak is consistent with the magnon Goldstone gap reported in previous INS and Raman results (*21*, *37*). It should be noted that the first peak-like feature is relatively weak, mixing with considerable background (Fig. 2C), which may be associated by other non-magnon related excitations in such an energy range.

To understand the magnetic-field dependent IETS and extract the physical parameters in our system, we start with the Hamiltonian as following to describe the spin excitations in single-layer CrI₃,

$$H = -J_1 \sum_{\langle mn \rangle} S_m \cdot S_n - J_2 \sum_{\langle\langle mn \rangle\rangle} S_m \cdot S_n - J_3 \sum_{\langle\langle\langle mn \rangle\rangle\rangle} S_m \cdot S_n + \sum_{\langle\langle mn \rangle\rangle} D_z \cdot (S_m \times S_n)$$

$$- K \sum_m (S_m^z)^2 - h_B \sum_m S_m^z,$$

where $\langle \rangle$, $\langle\langle \rangle\rangle$, $\langle\langle\langle \rangle\rangle\rangle$ denote the sum over the nearest, next-nearest and third-nearest neighbors (TNN), respectively, $K$ is the anisotropic interaction factor, and $h_B$ represents the external magnetic field. The DMIs originating from spin-orbit coupling open a gap at the Dirac points. The unpaired electrons in the *d*-orbitals of each $Cr^{3+}$ contribute a spin state of $S = 3/2$. We analytically fit the theoretical magnon bands to the experimental spin excitations, with $J_1 = 2.12\ meV$, $J_2 = 0.15 J_1$, $J_3 = -0.04 J_1$, $D_z = 0.08 J_1$ and $K = 0.12 J_1$ (*16*). Here, $J_1$ and $J_3$ are comparable to the bulk values, while $J_2$ and $D_z$ are found to be twice larger (*21*). Besides the DMIs associated with the lacking of inversion symmetry in the intralayer bonding (*10*), the interlayer superexchange coupling has considerable contribution to gap opening in the multilayer crystals (*12*, *21*). In the single-layer CrI₃ grown on Au(111), the interfacial Rashba SOC may play an important role on enhancing the DMIs. In addition, the Ruderman-Kittel-Kasuya-Yosida (RKKY) interaction on the Au substrate may also facilitate the long-range coupling and the magnetic anisotropy (*38–41*).

Since magnon-magnon interaction may occur at finite temperatures, we performed an effective renormalization via the application of the Zeeman field as the mean field (*16*). The best-fit values highlight the self-energy correction calling into the temperature dependence to avoid the breakdown of topological edge modes even containing magnon-magnon interactions. This field-dependent renormalization yields a nonlinear evolution of the Goldstone gap $\Delta_G = h_B S + 2KS - 2N_K S e^{-\frac{h_B}{T}}$ (*42*). Arising from the flat and the saddle-shaped band structure, two vHSs deviate from their zero-field values determined by the high-order corrections of Heisenberg exchange $N_J$. The theoretical results of the field dependence of the magnon DOS are shown in Fig. 2B and 2D, which grasp the main feature of our experiments.

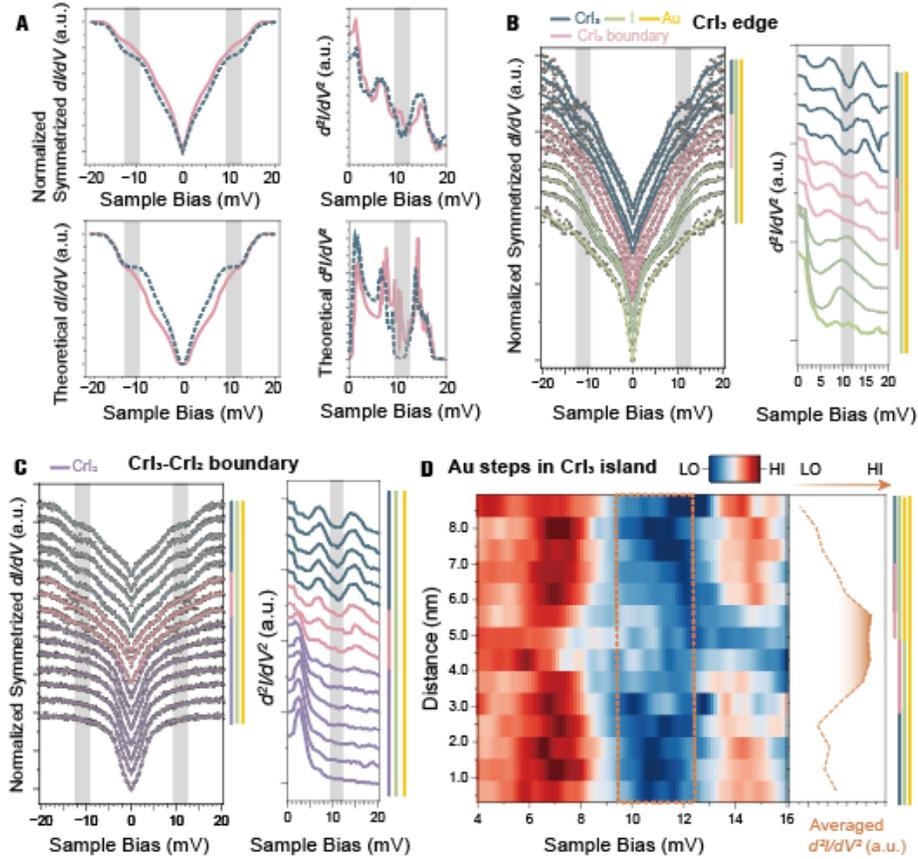

**Fig. 3. IETS at the edges of single-layer CrI₃.** (**A**) Typical *dI/dV* and *d²I/dV²* curves acquired from the edge (doted pink lines) compared with those from the interior (navi lines). Top: experimental IETS showing extra signals within the energy gap. Bottom: theoretically simulated curves based on the Heisenberg-DM model. (**B**) IETS results across the CrI₃ edge. Left: *dI/dV* curves are shown in circles and the smoothed results are shown by color lines. Right: *d²I/dV²* from the smoothed *dI/dV*. (**C**) IETS results across a CrI₃-CrI₂ boundary. (**D**) IETS results across a CrI₃ covered Au(111) step edge, at which there are two CrI₃ edges (from upper and lower terraces). Left: a color map showing the position-dependent intensity of *d²I/dV²*, indicating stronger in-gap signals from the step edge. Right: Averaged intensity in the energy range of 9.5-12.4 meV as a function of position. To the right of the panels in **B-D** are the diagrams of three types of samples, showing the STM tip positioning for each IETS measurement.

**Evidence for topological magnon edge states**

To identify the possible contribution of magnon edge states to the inelastic tunneling conductance, we conducted IETS measurements at the CrI₃ edges, in comparison with the data acquired from the interior of CrI₃ islands. Typical IETS results acquired from the edge and interior of CrI₃, as well as the theoretically simulated local magnon spectra of a single-layer CrI₃ ribbon are shown in Fig. 3A. One can find that, in the magnon gap, there are extra signal in the IETS from the edge. In the *dI/dV* channel, the plateau centered at 11.0 meV disappeared. Instead, there is an ascending slope, indicating the continuously increased tunneling conductivity. Meanwhile, in the *d²I/dV²* curve, extra signals were found in the gap, confirming the existence of magnon edge states. Such additional DOS of magnons can be well mimicked theoretically by considering an open-boundary CrI₃ ribbon (Fig. S11).

We further investigated the spatial dependence of the edge states. First, line-profile spectra across the island edge, which is the boundary between CrI$_3$ and the vacuum, were collected. As shown in Fig. 3B, the IETS from the iodine buffer layer exhibits a steeper increase of conductance below 2.0 meV and then a slow increase from 6.0 to 20.0 meV. In the $d^2I/dV^2$ channel, there is a plateau from 0 to 2.0 meV, following a featureless signal from 6.0 to 20.0 meV. Throughout all IETS curves, the plateau from 0 to 2.0 meV persists in $d^2I/dV^2$ channel, indicating that the previously discussed non-magnon related background may originate from the iodine layer on the substrate. As the STM tip positioning closer to the island, an extra $d^2I/dV^2$ peak located at about 9 meV appears. When positioning the tip upon the edge of the CrI$_3$ island, the obtained $d^2I/dV^2$ curves exhibit similar vHSs as those at the interior of the CrI$_3$ island, but filling with extra signals in the gap centered at 11.0 meV. When the tip was moved towards the interior of the island, the gap-like feature gradually became distinct, while the in-gap states were attenuated and finally disappeared.

The above results unveiled both the energy gap and the edge states of magnon spectra in 2D CrI$_3$ islands, agreeing well with the picture of the topological magnon states. We should also rule out other possible origins such as dangling bonds at the edges. For this reason, we prepared single-layer CrI$_3$-CrI$_2$ lateral junctions with atomically seamless boundaries (*31*, *43*). Since single-layer CrI$_2$ with a closely packed layer of chromium ions with trigonal lattice processes a topologically trivial magnon band structure, magnon edge states are expected to emerge at CrI$_3$-CrI$_2$ boundaries. As shown by the purple lines in Fig. 3C, the $d^2I/dV^2$ curves on CrI$_2$ exhibits a predominant peak at about 2.5 meV, while at the interior of the CrI$_3$ side, there are typical gap opening and vHSs as described above. At the boundary, we did observe the extra signals between the two vHSs.

Finally, we investigated the CrI$_3$ edges located at a step of the Au(111) surface, where two CrI$_3$ edges from upper and lower terraces join together but without any chemical bonding in between. As shown in Fig. 3D, extra contributions from the edge states can be clearly observed near the Au(111) step, the detectable signals of which are spatially distributed in a region with the width about 6.0 nm. At the same time, the topological gap can be always observed on both sides of the step edges. Therefore, we conclude that the extra signals observed in the magnon gap is not ascribed to the average effect, which mixes the magnon spectra from both sides of a CrI$_3$ edge, resulting in averaged DOS at the edge. In addition, the persistence of edge states at the Au(111) steps indicates the negligible coupling between the two magnon systems from the upper and lower terraces, though the height of the gold step (2.4 Å) is much smaller than the thickness of single-layer CrI$_3$(6.9 Å).

In summary, by investigating the magnon spectra at three different types of boundaries, possible trivial origins of the boundary states have been ruled out, verifying the emergence of topologically protected non-trivial magnon edge states of single-layer CrI$_3$ as a 2D MCI. Our experiments revealed the highly-confined nature of the spatial distribution of the magnon edge states, with the DOS dominantly located at the edges with a width of several nanometers. The localized feature makes it difficult to observe by bulk-sensitive spectroscopic techniques. Although thermally excited magnons contribute to the depletion of magnetization, our results highlight the topological characteristics in MCIs against perturbations and reductions of the spin stiffness. The localized chiral edge states can withstand various defects and impurities, maintaining their integrity to support robust propagation in waveguides and other spintronic devices (*44*).

**Discussion and outlook**

In this work, we have directly observed the topological gap and edge states in the magnon spectra of single-layer CrI$_3$, verifying the existence of MCI states in such a 2D ferromagnet down to single-layer limit. There is a Zeeman-like energy shift of the magnon spectra induced by an out-of-plane external magnetic field, exhibiting a slightly nonlinear relationship which is attributed to the renormalization at finite temperatures. With a Heisenberg-DM Hamiltonian model of the 2D ferromagnetic honeycomb system including NN, NNN, and TNN superexchange interactions, we achieved an optimal fitting to the magnon spectra acquired experimentally. Much stronger NNN and Dzyaloshinskii-Moriya interactions were deduced in the 2D system compared with the multilayer counterpart. In case that MCIs are induced by DMIs with non-zero in-plane components, a topological phase transition driven by fine-tuned vector magnetic field is expected (*45*), which is possible to be further verified by using IETS.

Our study provides the first direct evidence of topologically protected edge states in MCIs, demonstrating that IETS based on STM is a powerful technique for spatially resolved detection of localized spin excitations. Our results show the robustness of the magnon edge states against magnon-magnon interactions, which may result in the breakdown of edge modes (*46*), implying the validation of linear spin-wave theory. The chiral edge modes with 1D nature is advantageous for the development of dissipationless spintronic devices (*4*). In particular, the $CrI_2$-$CrI_3$ junctions holding topologically protected magnon boundary states can serve as a research platform for MCIs and facilitate the construction for future magnonic devices.

**References and Notes**

1. X. S. Wang, X. R. Wang, Topological magnonics. *Journal of Applied Physics* **129**, 151101 (2021).

2. P. A. McClarty, Topological Magnons: A Review. *Annu. Rev. Condens. Matter Phys.* **13**, 171–190 (2022).

3. B. Lenk, H. Ulrichs, F. Garbs, M. Münzenberg, The building blocks of magnonics. *Physics Reports* **507**, 107–136 (2011).

4. A. V. Chumak, V. I. Vasyuchka, A. A. Serga, B. Hillebrands, Magnon spintronics. *Nature Phys* **11**, 453–461 (2015).

5. K. A. van Hoogdalem, Y. Tserkovnyak, D. Loss, Magnetic texture-induced thermal Hall effects. *Phys. Rev. B* **87**, 024402 (2013).

6. L. Zhang, J. Ren, J.-S. Wang, B. Li, Topological magnon insulator in insulating ferromagnet. *Phys. Rev. B* **87**, 144101 (2013).

7. Y. Onose, T. Ideue, H. Katsura, Y. Shiomi, N. Nagaosa, Y. Tokura, Observation of the Magnon Hall Effect. *Science* **329**, 297–299 (2010).

8. M.-H. Zhang, D.-X. Yao, Topological magnons on the triangular kagome lattice. *Phys. Rev. B* **107**, 024408 (2023).

9. I. Dzyaloshinsky, A thermodynamic theory of "weak" ferromagnetism of antiferromagnetics. *Journal of Physics and Chemistry of Solids* **4**, 241–255 (1958).

10. T. Moriya, Anisotropic Superexchange Interaction and Weak Ferromagnetism. *Phys. Rev.* **120**, 91–98 (1960).

11. R. Chisnell, J. S. Helton, D. E. Freedman, D. K. Singh, R. I. Bewley, D. G. Nocera, Y. S. Lee, Topological Magnon Bands in a Kagome Lattice Ferromagnet. *Phys. Rev. Lett.* **115**, 147201 (2015).

12. L. Chen, J.-H. Chung, B. Gao, T. Chen, M. B. Stone, A. I. Kolesnikov, Q. Huang, P. Dai, Topological Spin Excitations in Honeycomb Ferromagnet $CrI_3$. *Phys. Rev. X* **8**, 041028 (2018).

13. S. A. Owerre, A first theoretical realization of honeycomb topological magnon insulator. *J. Phys.: Condens. Matter* **28**, 386001 (2016).

14. S. K. Kim, H. Ochoa, R. Zarzuela, Y. Tserkovnyak, Realization of the Haldane-Kane-Mele Model in a System of Localized Spins. *Phys. Rev. Lett.* **117**, 227201 (2016).

15. S. S. Pershoguba, S. Banerjee, J. C. Lashley, J. Park, H. Ågren, G. Aeppli, A. V. Balatsky, Dirac Magnons in Honeycomb Ferromagnets. *Phys. Rev. X* **8**, 011010 (2018).

16. see Supplementary Materials.

17. T. Ideue, Y. Onose, H. Katsura, Y. Shiomi, S. Ishiwata, N. Nagaosa, Y. Tokura, Effect of lattice geometry on magnon Hall effect in ferromagnetic insulators. *Phys. Rev. B* **85**, 134411 (2012).

18. M. A. McGuire, H. Dixit, V. R. Cooper, B. C. Sales, Coupling of Crystal Structure and Magnetism in the Layered, Ferromagnetic Insulator $CrI_3$. *Chem. Mater.* **27**, 612–620 (2015).

19. B. Huang, G. Clark, E. Navarro-Moratalla, D. R. Klein, R. Cheng, K. L. Seyler, D. Zhong, E. Schmidgall, M. A. McGuire, D. H. Cobden, W. Yao, D. Xiao, P. Jarillo-Herrero, X. Xu, Layer-dependent ferromagnetism in a van der Waals crystal down to the monolayer limit. *Nature* **546**, 270–273 (2017).

20. E. Aguilera, R. Jaeschke-Ubiergo, N. Vidal-Silva, L. E. F. F. Torres, A. S. Nunez, Topological magnonics in the two-dimensional van der Waals magnet $CrI_3$. *Phys. Rev. B* **102**, 024409 (2020).

21. L. Chen, J.-H. Chung, M. B. Stone, A. I. Kolesnikov, B. Winn, V. O. Garlea, D. L. Abernathy, B. Gao, M. Augustin, E. J. G. Santos, P. Dai, Magnetic Field Effect on Topological Spin Excitations in $CrI_3$. *Phys. Rev. X* **11**, 031047 (2021).

22. B. C. Stipe, M. A. Rezaei, W. Ho, Single-Molecule Vibrational Spectroscopy and Microscopy. *Science* **280**, 1732–1735 (1998).

23. C. F. Hirjibehedin, C.-Y. Lin, A. F. Otte, M. Ternes, C. P. Lutz, B. A. Jones, A. J. Heinrich, Large Magnetic Anisotropy of a Single Atomic Spin Embedded in a Surface Molecular Network. *Science* **317**, 1199–1203 (2007).

24. C. F. Hirjibehedin, C. P. Lutz, A. J. Heinrich, Spin Coupling in Engineered Atomic Structures. *Science* **312**, 1021–1024 (2006).

25. A. J. Heinrich, J. A. Gupta, C. P. Lutz, D. M. Eigler, Single-Atom Spin-Flip Spectroscopy. *Science* **306**, 466–469 (2004).

26. X. Chen, Y.-S. Fu, S.-H. Ji, T. Zhang, P. Cheng, X.-C. Ma, X.-L. Zou, W.-H. Duan, J.-F. Jia, Q.-K. Xue, Probing Superexchange Interaction in Molecular Magnets by Spin-Flip Spectroscopy and Microscopy. *Phys. Rev. Lett.* **101**, 197208 (2008).

27. A. Mook, J. Henk, I. Mertig, Edge states in topological magnon insulators. *Phys. Rev. B* **90**, 024412 (2014).

28. L. Vitali, M. A. Schneider, K. Kern, L. Wirtz, A. Rubio, Phonon and plasmon excitation in inelastic electron tunneling spectroscopy of graphite. *Phys. Rev. B* **69**, 121414 (2004).

29. S. C. Ganguli, M. Aapro, S. Kezilebieke, M. Amini, J. L. Lado, P. Liljeroth, Visualization of Moiré Magnons in Monolayer Ferromagnet. *Nano Lett.* **23**, 3412–3417 (2023).

30. J. Feldmeier, W. Natori, M. Knap, J. Knolle, Local probes for charge-neutral edge states in two-dimensional quantum magnets. *Phys. Rev. B* **102**, 134423 (2020).

31. P. Li, C. Wang, J. Zhang, S. Chen, D. Guo, W. Ji, D. Zhong, Single-layer $CrI_3$ grown by molecular beam epitaxy. *Science Bulletin* **65**, 1064–1071 (2020).

32. D. R. Klein, D. MacNeill, J. L. Lado, D. Soriano, E. Navarro-Moratalla, K. Watanabe, T. Taniguchi, S. Manni, P. Canfield, J. Fernández-Rossier, P. Jarillo-Herrero, Probing magnetism in 2D van der Waals crystalline insulators via electron tunneling. *Science* **360**, 1218–1222 (2018).

33. L. Webster, L. Liang, J.-A. Yan, Distinct spin–lattice and spin–phonon interactions in monolayer magnetic $CrI_3$. *Phys. Chem. Chem. Phys.* **20**, 23546–23555 (2018).

34. P. Delugas, O. Baseggio, I. Timrov, S. Baroni, T. Gorni, Magnon-phonon interactions enhance the gap at the Dirac point in the spin-wave spectra of $CrI_3$ two-dimensional magnets. *Phys. Rev. B* **107**, 214452 (2023).

35. M.-H. Zhang, D.-X. Yao, Type-II Dirac points and Dirac nodal loops on the magnons of a square-hexagon-octagon lattice. *Phys. Rev. B* **108**, 144407 (2023).

36. L. Ke, M. I. Katsnelson, Electron correlation effects on exchange interactions and spin excitations in 2D van der Waals materials. *npj Comput Mater* **7**, 4 (2021).

37. J. Cenker, B. Huang, N. Suri, P. Thijssen, A. Miller, T. Song, T. Taniguchi, K. Watanabe, M. A. McGuire, D. Xiao, X. Xu, Direct observation of two-dimensional magnons in atomically thin $CrI_3$. *Nat. Phys.* **17**, 20–25 (2021).

38. M. A. Ruderman, C. Kittel, Indirect Exchange Coupling of Nuclear Magnetic Moments by Conduction Electrons. *Phys. Rev.* **96**, 99–102 (1954).

39. T. Kasuya, A Theory of Metallic Ferro- and Antiferromagnetism on Zener's Model. *Prog. Theor. Phys.* **16**, 45–57 (1956).

40. K. Yosida, Magnetic Properties of Cu-Mn Alloys. *Phys. Rev.* **106**, 893–898 (1957).

41. J.-J. Zhu, D.-X. Yao, S.-C. Zhang, K. Chang, Electrically Controllable Surface Magnetism on the Surface of Topological Insulators. *Phys. Rev. Lett.* **106**, 097201 (2011).

42. M. Bloch, Magnon Renormalization in Ferromagnets Near the Curie Point. *Phys. Rev. Lett.* **9**, 286–287 (1962).

43. P. Li, N. Liu, J. Zhang, S. Chen, X. Zhou, D. Guo, C. Wang, W. Ji, D. Zhong, Two-Dimensional Magnetic Semiconducting Heterostructures of Single-Layer $CrI_3$–$CrI_2$. *ACS Appl. Mater. Interfaces* **15**, 19574–19581 (2023).

44. J. Li, T. Datta, D.-X. Yao, Einstein-de Haas effect of topological magnons. *Phys. Rev. Research* **3**, 023248 (2021).

45. V. Brehm, P. Sobieszczyk, J. N. Kløgetvedt, R. F. L. Evans, E. J. G. Santos, A. Qaiumzadeh, Topological magnon gap engineering in van der Waals $CrI_3$ ferromagnets. *Phys. Rev. B* **109**, 174425 (2024).

46. J. Habel, A. Mook, J. Willsher, J. Knolle, Breakdown of chiral edge modes in topological magnon insulators. *Phys. Rev. B* **109**, 024441 (2024).

47. C. J. Chen, *Introduction to Scanning Tunneling Microscopy* (Oxford University Press, ed. 3, 2021).